\documentclass[sigconf]{acmart}

\usepackage{booktabs} 

\usepackage[utf8]{inputenc}

\usepackage{subcaption}

\usepackage{tabularx}

\usepackage[]{csquotes}
\renewcommand{\mkbegdispquote}[2]{\itshape}

\usepackage{enumitem}

\usepackage{needspace}

\usepackage{balance}

\usepackage{dcolumn}
\newcolumntype{d}[1]{D..{#1}} 

\copyrightyear{2018} 
\acmYear{2018} 
\acmConference[LBRS@RecSys '18]{Late-Breaking Results 
track part of the Twelfth ACM Conference on Recommender Systems}{October 2--7, 2018}{Vancouver, BC, Canada}
\acmBooktitle{Late-Breaking Results 
track part of the Twelfth ACM Conference on Recommender Systems (RecSys '18), October 2--7, 2018, Vancouver, BC, Canada}
\acmISBN{} 
\acmDOI{} 

\setcopyright{rightsretained}

\begin{document}

\title{Early Lessons from a Voice-Only Interface for Finding Movies}

\author{Joshua Wissbroecker}
\affiliation{\institution{University of Minnesota}}
\email{wissb004@umn.edu}

\author{F. Maxwell Harper}
\orcid{0000-0003-0552-5773}
\affiliation{\institution{University of Minnesota}}
\email{max@umn.edu}




\author{}


\begin{abstract}
The current generation of streaming media players often allow users to speak commands (e.g., users can change the TV channel by pressing a button and saying ``ESPN''). However, these devices typically support a narrow range of control- and search-oriented commands, and do not support deeper recommendation or exploration queries. To study spoken natural language interactions with recommenders, we have built MovieLens TV, a movie recommender system with no input modalities except voice. In this poster, we describe MovieLens TV, with a focus on lessons learned building a prototype system around an off-the-shelf Amazon Echo.
\end{abstract}

%
%



\maketitle





\section{Introduction}

As voice recognition continues to improve, voice user interfaces are becoming more capable and pervasive. As recommender systems researchers, it is exciting to consider the potential for rich user interactions from voice input. For example, recent work~\cite{kang_understanding_2017} found that users issue a broad range of queries when interacting with a natural language recommender. However, the current generation of voice-controlled streaming devices emphasize a small set of command- and search-oriented commands over a deeper recommendation experience.

To study voice user interfaces for recommendation, we built a prototype system called MovieLens TV using web technologies and an Amazon Echo. This system offers no input modalities except for speech. It presents responses on-screen and using voice output; it incorporates a recommendation engine to personalize the display of results. This poster paper describes the architecture of MovieLens TV and several challenges to developing voice user interfaces for recommender systems using current off-the-shelf hardware.

\begin{figure}[t]
  \centering

  \begin{subfigure}[b]{\columnwidth}
    \includegraphics[width=\columnwidth]{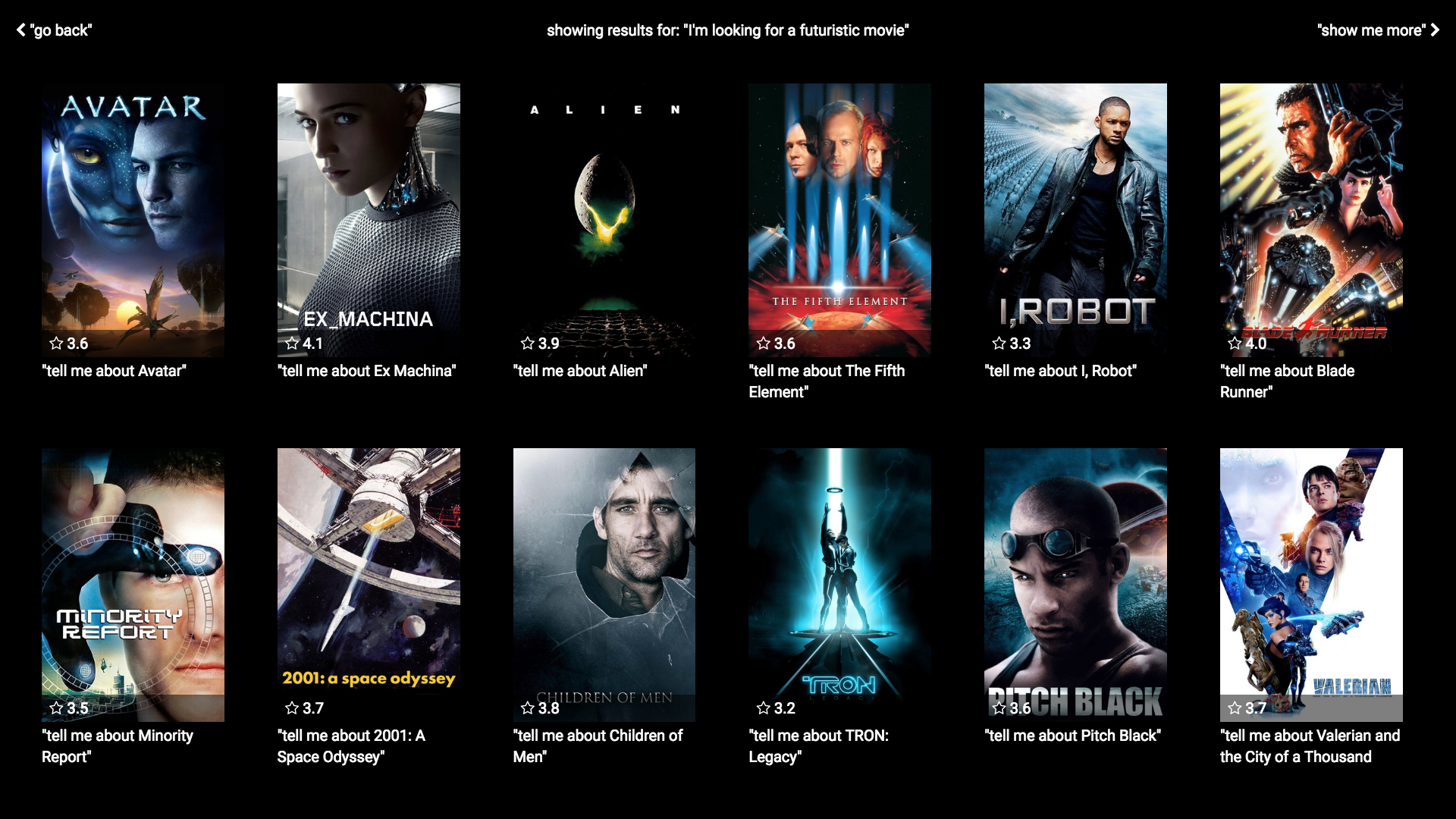}
    \label{fig:screenshots-top}
  \end{subfigure}

  \begin{subfigure}[b]{\columnwidth}
    \includegraphics[width=\columnwidth]{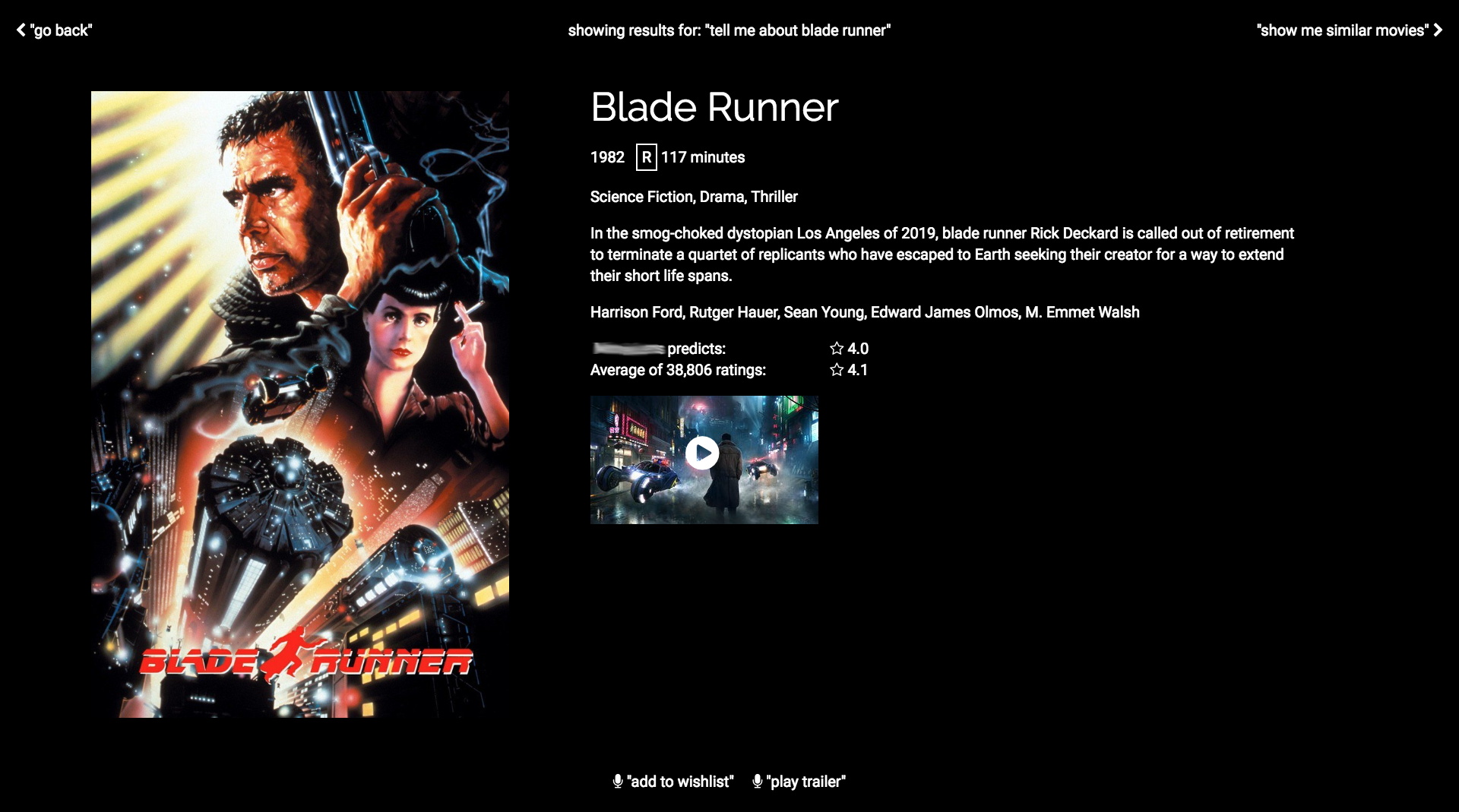}
  \end{subfigure}

  \caption{Screenshots of the MovieLens TV interfaces: the explore view (top) and the details view (bottom). Because there are no available input modalities except for voice, we label available actions in the interface in quotes (e.g., ``go back'').}
  \label{fig:screenshots}
\end{figure}

\section{MovieLens TV System}


MovieLens TV is a prototype system that requires no custom hardware --- it can be run with an Amazon Echo and a web browser. To use the system, a user must install and sign in to our custom Alexa skill on the Echo. After this one-time setup, the user opens the app in a web browser and says ``Alexa, open MovieLens'' to launch the voice interface. At this point, the user may begin sending requests, prefixed by ``Alexa'', to ensure that the Echo is listening. For example, the user might say ``Alexa, show action movies'', followed by ``Alexa, show me more''.

\begin{figure}[t]
  \centering
  \includegraphics[width=0.4\columnwidth]{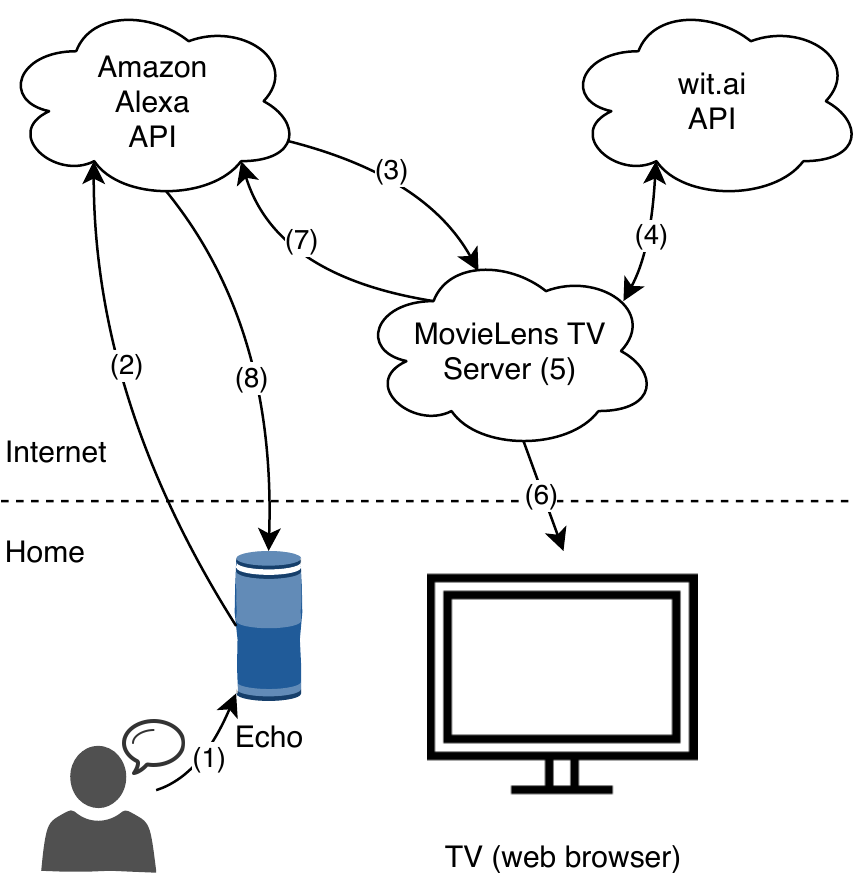}
  \caption{MovieLens TV system architecture. (1) The user speaks a request to the Echo (``Alexa, I'm looking for futuristic movies''). (2) The Echo sends the audio to Amazon's Alexa API, where we have created a skill. (3) Amazon sends the transcribed text of the query to our server as unstructured text. (4) Our server sends the text to \href{https://wit.ai/}{wit.ai}, which returns a structured representation of the text, including an intent and other extracted features. (5) Our server uses the structured representation of the request to determine an action to take, then retrieves the relevant information, such as recommendations or movie information. (6) Our server pushes a message to the browser using a WebSocket, and the browser updates the view. (7) Our server sends a text response to Amazon. (8) The Echo vocalizes the text response. (6) and (8) are approximately simultaneous.}
  \label{fig:flowchart}
\end{figure}

For user testing, we only allow users to drive the user interface by voice (i.e., we do not allow the user to navigate using a mouse, keyboard, or remote control). To achieve this, we developed the architecture shown in Figure~\ref{fig:flowchart}. Central to this architecture is the idea that the MovieLens TV server pushes WebSocket messages to the client, which uses JavaScript to render views. The MovieLens server internally maintains the state of each user, supporting queries like ``play the trailer'' (which requires knowledge of the currently-displayed movie), or ``show more'' (which requires knowledge of the previous search).

MovieLens TV has several views (see Figure~\ref{fig:screenshots} for two examples), including an \textit{explore} view for displaying movies in a grid, a \textit{details} view for viewing information about a movie, and a home/help screen. Simultaneous with showing the view, the Echo vocalizes a response; we are currently experimenting with design options. For early testing, we vocalize brief summaries of the action (e.g., ``Here are some movies that I think are futuristic'').

The \textit{explore} view supports top-N requests (e.g., ``show me science fiction movies'', or ``what are some popular comedies?''), as well as related-item recommendation requests (``show me more like Pitch Black''). We model these two types of requests as different intents. For top-N requests, unless we detect an explicit sort order (e.g., ``recent'', or ``popular''), we rank results using the MovieLens item-based K-nearest neighbors collaborative filtering algorithm. For related-item requests, we use a content-based method to rank the candidates, most similar movies first. For user testing, we instruct users to first create a personalized profile using the cold start interface in MovieLens.

\section{Challenges and Lessons}

\textbf{Voice triggers.} The Echo has two important limitations for researchers wishing to develop voice-driven systems. First, it requires that each request is prefixed by an activation word (by default, ``Alexa''). In early testing, we have found that this interrupts users' natural flow, and is a barrier for users to engage in a two-way dialogue. Second, since the Echo supports many commands (not just those related to MovieLens), the user must either constantly re-open the app, or further prefix queries (``Alexa, ask MovieLens to show action movies''), further draining the dialogue of its natural qualities. We address this second limitation with a widely-used hack: we append a silent audio clip to the end of each response, forcing the connection to stay open. Unfortunately, this method times out after two minutes, and if the user has not issued a new query during this window, he or she must again say ``Alexa, open MovieLens''. Ultimately, these challenges make the user experience much more ``request/response'' than ``conversational'', a limitation to current voice interfaces identified and discussed in depth in \cite{porcheron_voice_2018}.


\textbf{Intent detection.} Alexa skills are typically built around Amazon's own intent detection process, which is a way of mapping several ways of asking for something (a goal) to a structured representation including any variables (i.e., \textit{slots}) present in the query~\cite{xu_convolutional_2013}. However, Amazon's method relies on pattern matching, which is not flexible enough to accommodate the broad range of queries that early testing revealed. We therefore use a third party intent detection service (\href{https://wit.ai/}{wit.ai}), which uses supervised machine-learning to predict user intent. Using this approach has several advantages: its accuracy improves as we train more example queries, and it is often able to correctly guess the intent of a query that we have not seen before. However, we have found that the intent detection approach has some inherent limitations, such as disambiguating between and/or, and determining when the user is attempting to build on the current set of results rather than starting a fresh query. Mapping natural language to recommendation tasks using intent detection or other approaches is an open challenge for researchers.

\section{Next Steps}

In early user testing, we have found that MovieLens TV is a minimum viable product with many areas for improvement. Users are not familiar with the voice-only input modality, and find it helpful to see example queries on the screen. These hints, although they assist in usability, may limit users' creativity in formulating their own requests. Users are surprised at the range of queries they can express. Equally surprising, however, are the many cases where the system can break, often due to transcription errors: it is still difficult to accurately transcribe some actors' names or movie titles.

We are currently conducting a user study to learn more about the possibilities and challenges involved with developing voice user interfaces for recommender systems. We are excited to learn more about users' expectations concerning voice-based recommendations, to document the most prominent failure cases, and to continue working towards more effective recommendation technology.

\section{Acknowledgments}

This material is based on work supported by the National Science Foundation under grant IIS-1319382, by a grant from Amazon, and by the University of Minnesota’s Undergraduate Research Opportunities Program.


\bibliographystyle{ACM-Reference-Format}
\bibliography{movietv-poster} 

\end{document}